# Direct Access to 5G New Radio User Equipment from NGSO Satellites in Millimeter Waves

Pantelis-Daniel Arapoglou
*Radio Frequency Systems Division*
*ESA/ESTEC*
Noordwijk, The Netherlands
pantelis-daniel.arapoglou@esa.int

Stefano Cioni
*Radio Frequency Systems Division*
*ESA/ESTEC*
Noordwijk, The Netherlands
Stefano.Cioni@esa.int

Emiliano Re
*Radio Frequency Systems Division*
*ESA/ESTEC*
Noordwijk, The Netherlands
Emiliano.Re@esa.int

Alberto Ginesi
*Radio Frequency Systems Division*
*ESA/ESTEC*
Noordwijk, The Netherlands
Alberto.Ginesi@esa.int

*Abstract*—The next generation of terrestrial radio communications, so-called fifth generation (5G) New Radio (NR), beyond the traditional bands below 6 GHz, has been also specified to operate over millimeter waves (mmWaves), in the so-called Frequency Range 2 (FR2). Such frequency bands have been since decades the 'natural habitat' for fixed satellite services (FSS). In this new landscape, this paper preliminary investigates the feasibility of non-geostationary orbit (NGSO) satellites directly accessing NR-enabled User Equipment (UE) in mmWaves, from a regulatory, UE characteristics, space segment, link budget and system point of view. It also identifies future R&D needs in this area.

*Keywords—5G, New Radio, millimeter waves, non-geostationary orbit, user equipment.*

## I. INTRODUCTION

As part of the International Mobile Telecommunications (IMT)-2020 vision, the next generation of terrestrial radio communications is being developed under the third generation partnership project (3GPP), the so-called *5G New Radio* (NR). In this new generation of terrestrial radio, aside of the conventional cellular radio bands below 6 GHz, also frequency bands above 6 GHz will be employed in the millimeter wave (mmWave) range for the first time with applications for short range or indoor links [1].

On the other hand, mmWaves have been since decades the 'natural habitat' of fixed satellite services (FSS), typically in the geostationary orbit (GEO). An additional development in the space sector has been the launch of at least two and the announcement of many more large satellite constellations in non-geostationary orbit (NGSO), typically Low Earth Orbit (LEO) [2].

In this new landscape, and in pursuing the maximum integration of satellite networks into 5G, one comes naturally to ask whether broadband direct access from NGSO satellites to low gain (terrestrial-grade) handheld User Equipment (UE) is feasible. In addition, an interesting question is whether the terrestrial and satellite access can share the same frequency bands and provide complementary coverage based on power levels. It is also noted that typically the bands above 10 GHz are meant to be employed only with highly directional (dish-based) satellite terminals for reasons related to both power efficiency as well as interference avoidance. The paper attempts to carry out a preliminary feasibility analysis by investigating regulatory aspects (particularly spectrum usage) in Section II, UE characteristics in mmWave in Section III, space segment design in Section V, link budget and system aspects in Sections IV and VI. The intention of it is not to provide a final answer to this feasibility question but rather to stimulate further research on the topic and identify technology limitations and future R&D needs.

## II. REGULATORY ASPECTS

This section briefly discusses the regulatory situation from a spectrum management point of view and derives conclusions at the system level.

### A. 5G NR Frequency Bands

Frequency bands for 5G NR are separated into two different frequency ranges. First, there is *Frequency Range 1* (FR1) that includes sub-6GHz frequency bands, some of which are bands traditionally used by previous standards, but have been extended to cover potential new spectrum. The other is *Frequency Range 2* (FR2) [3] that includes frequency bands from 24.25 GHz to 52.6 GHz. The bands in this mmWave range have shorter range but higher available bandwidth than the bands in the FR1. The bands allocated to FR2 5G NR are listed in TABLE I.

TABLE I.    mmWAVE BANDS ALLOCATE TO FR2 5G NR.

| NR band no. | Freq. band | Uplink/Downlink (MHz) | Bandwidth |
|---|---|---|---|
| n257 | 28 GHz | 26500-29500 | 3 GHz |
| n258 | 26 GHz | 24250-27500 | 3.25 GHz |
| n260 | 39 GHz | 37000-40000 | 3 GHz |
| n261 | 28 GHz | 27500-28350 | 850 MHz |

One important element to highlight with respect to TABLE I. is that the duplexing scheme foreseen is Time Division Duplexing (TDD). That is, both uplink and downlink operations take place over the same band and are separated only in the time domain[1]. Another element to stress is the fact that, as Fig. 1 shows, the Ka-band part of the 5G

---

[1] In other words, only half-duplex communications are allowed, thus the device cannot transmit and receive at the same time.

FR2 allocation, differs significantly among the various regions of the globe [4].

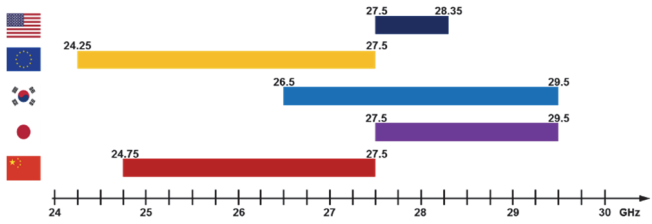

Fig. 1. Regional allocation of spectrum for 5G FR2 (Ka-band) [4].

*B. Shared Bands between Terrestrial and Space Services*

The bands presented in the previous sub-section are from a 3GPP point of view. From an International Telecommunications Union- Radiocommunications sector (ITU-R) point of view, these bands are shared, i.e. they are allocated on a primary basis to the (terrestrial) Fixed Service (FS), the (terrestrial) Mobile Service (MS), and the FSS. TABLE II. presents this status for typical sub-bands around 28 GHz and 39 GHz based on the ITU-R Radio Regulation table of Frequency Allocations.

TABLE II. ITU-R RADIO REGULATIONS FREQUENCY ALLOCATION STATUS FOR TYPICAL SUB-BANDS AROUND 28 AND 39 GHZ.

| | |
|---|---|
| 27.5-28.5 | FIXED 5.537A<br>FIXED-SATELLITE (Earth-to-space) 5.484A 5.516B 5.539<br>MOBILE<br>5.538 5.540 |
| 38-39.5 | FIXED<br>FIXED-SATELLITE (space-to-Earth)<br>MOBILE<br>Earth exploration-satellite (space-to-Earth)<br>5.547 |

It is interesting to point out that the bands allocated to a satellite service come with a clear indication of whether the band should be used from Earth-space (uplink) or from space-Earth (downlink). Such a constraint does not exist for the bands allocated to terrestrial services. This fact, combined with the TDD approach that has been adopted by 5G networks, leads to the important conclusion that *5G NR networks and FSS networks share a downlink band only in 39 GHz (Q-band) and an uplink band only in the 28 GHz*.

The boundaries for sharing the bands that are allocated to both satellite and terrestrial services are described in Article 21 of the ITU-R Radio Regulations. The constraint put on the satellite service to protect the terrestrial services comes in the form of a Power Flux Density (PFD) limit over a reference bandwidth [dBW/m$^2$ over 1 MHz]. This limit depends on the elevation angle, as lower elevation angles represent increased risk of interference into terrestrial systems. The PFD limits for the bands of interest are depicted in TABLE III.

Concluding this section on regulatory aspects, in the Ka downlink band, in addition to the shared band, there exists also an exclusive satellite band of 500 MHz between 19.7 GHz and 20.2 GHz. Exclusive implies that it is only allocated to satellite services (FSS) and not to terrestrial services. This band, which obeys to different constraints and limitations, is outside the scope of the present preliminary analysis. Further, in Q-band downlink (39 GHz), there is no similar exclusive satellite band. Therefore, the rest of this document discusses exclusively the shared bands discussed in this section.

TABLE III. PFD LIMITS OVER REFERENCE BANDWIDTH [DBW/M$^2$/MHZ] FOR SHARED BANDS AROUND 19 GHZ AND 39 GHZ.

| Frequency band | Service* | Limit in dB(W/m$^2$) for angles of arrival ($\delta$) above the horizontal plane | | | Reference bandwidth |
|---|---|---|---|---|---|
| | | 0°-5° | 5°-25° | 25°-90° | |
| 17.7-19.3 GHz [7,8] | Fixed-satellite (space-to-Earth) | −115 [13, 13A]<br>or | −115 + 0.5($\delta$ − 5) [13, 13A]<br>or | −105 [13, 13A]<br>or | 1 MHz |
| 19.3-19.7 GHz | Fixed-satellite (space-to-Earth) | −115 [13A] | −115 + 0.5($\delta$ − 5) [13A] | −105 [13A] | 1 MHz |
| 37.5-40 GHz | Fixed-satellite (non-geostationary-satellite orbit)<br>Mobile-satellite (non-geostationary-satellite orbit) | −120 [10, 16] | −120 + 0.75($\delta$ − 5) [10,16] | −105 [10, 16] | 1 MHz |
| 37.5-40 GHz | Fixed-satellite (geostationary-satellite orbit)<br>Mobile-satellite (geostationary-satellite orbit) | 0°-5°<br>−127 [16] | 5°-20°<br>−127 + (4/3)($\delta$ − 5) [16] | 20°-25°  25°-90°<br>−107 + 0.4  −105 [16]<br>($\delta$ − 20) [16] | 1 MHz |

### III. USER EQUIPMENT CHARACTERISTICS

To allow for a practical use of the mmWave range, the communication paradigm of 5G NR FR2 has changed from being cell-centric to beam-centric [5]. This translates into base stations (BSs) as well as UEs communicating via directional beams that are formed through hybrid analog/digital beamforming by antenna arrays. This allows for some antenna directionality to balance out the high link losses, but also changes significantly the UE radio frequency (RF) specifications. The 3GPP standardization group in charge of such RF aspects is mainly Radio Access Network (RAN)-4.

The key handheld UE elements at 28 GHz and 39 GHz, listed in TABLE IV. and TABLE V. , respectively, are collected from various sources [6]-[10]. It is important to underline that some parameters are fixed or determined a-priori, due to regulatory aspects, like the peak Effective Isotropically Radiated Power (EIRP) upper limit specified by the Federal Communications Commission (FCC), or due to the terminal form factor, like the size of the phased-array antenna. On the other hand, some RF characteristics are presented in TABLE IV. and TABLE V. in a range, since the precise values are dependent on the chosen technology and the final UE product manufacturing. For example, the main contributions to the line implementation loss are due to RF coupling loss, pointing and beamforming errors, and any other source related to miniaturization and assembling issues.

The nomenclature ($M,N,P,M_g,N_g$) refers to a rectangular antenna array of $M$x$N$ elements in P polarizations (antenna panel) that is repeated on a grid of $M_g$x$N_g$ across the area of the UE. This means that rectangular shaped antenna panels are placed in various points of the UE to avoid blockage when the device is held by its user [11], [12].

TABLE IV. KEY UE TRANSMIT AND RECEIVE CHARACTERISTICS AT 28 GHZ [6]-[10].

| UE Parameter | Value |
| --- | --- |
| Antenna gain per element | 5 – 8 dBi |
| (Nominal) Peak EIRP | 25 dBm |
| Peak EIRP allowed by FCC | 43 dBm |
| Noise figure | 5-10 dB |
| Noise PSD | -169 : - 164 dBm/Hz |
| Implementation loss | 3.3-10 dB |
| Typical array size | 4x4, 8x8, 16x16 $(M, N, P, M_g, N_g) = (2, 4, 2, 1, 2)$; |
| Distance between array elements | $\lambda/2$ |

TABLE V. KEY UE TRANSMIT AND RECEIVE CHARACTERISTICS AT 39 GHZ [6]-[10].

| UE Parameter | Value |
| --- | --- |
| Antenna gain per element | 5 – 8 dBi |
| (Nominal) Peak EIRP | 23 dBm |
| Peak EIRP allowed by FCC | 43 dBm |
| Noise figure | 5-10 dB |
| Noise PSD | -169 : - 164 dBm/Hz |
| Implementation loss | 5-12 dB |
| Typical array size | 4x4, 8x8, 16x16 $(M, N, P, M_g, N_g) = (2, 4, 2, 1, 2)$; |
| Distance between array elements | $\lambda/2$ |

## IV. INSIGHTS ON LINK BUDGET

In Section II it was explained that 5G NR and FSS only share the (satellite downlink) band at 39 GHz and, partially, the (satellite uplink) band at 28 GHz in some regions of the globe. Nevertheless, to attain a deeper insight on the system performance, in this section more frequency bands are considered for the downlink budgets in the 19/39/73 GHz as well as the uplink budgets in the 28/50/83 GHz[2]; these include of course all the main downlink and uplink FSS bands beyond 10 GHz.

### A. Channel Modeling

A fundamental element of any wireless system operating in mmWaves is the effect the atmosphere has on radiowave propagation. At frequency bands above 10 GHz, a satellite link is degraded by rain, cloud and gaseous attenuation, as well as tropospheric scintillations [13]. This total attenuation is plotted in Fig. 2 versus elevation for a hypothetical link in Berlin employing the predication models in [14][3]. The figure is plotted for the three carrier frequencies of interest and corresponds to a specific link availability that is probably more favourable than typically required. Despite this, it is clear that a NGSO satellite that will communicate with the UE under low elevation angles for long periods of time will be particularly hit hard by the atmosphere.

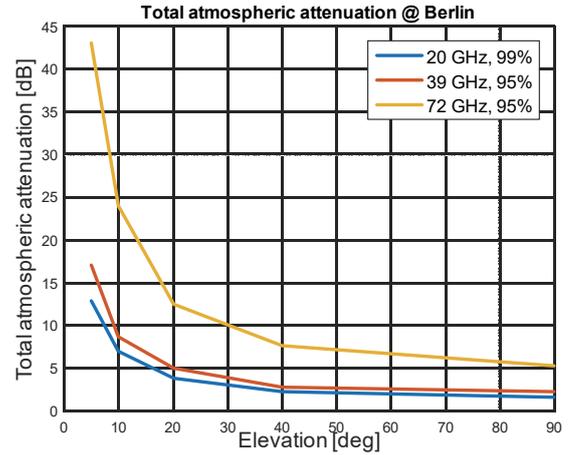

Fig. 2. Total atmospheric attenuation versus elevation for a hypothetical link in Berlin employing the prediction models in [14].

However, it is expected that a constellation of LEO satellites providing direct access to handheld terminals would be quite dense and thus each user would be served at relatively high elevation angle Nevertheless, in the downlink one could also envisage a downlink power control scheme, where the satellite adjusts its output power depending on the atmospheric conditions. In such a case, if the on board power allows, a constant PFD on ground approaching the allowed ITU-R limit can be maintained. Unlikely, a similar uplink power control scheme does not seem possible in the uplink from the power limited UE, as the peak EIRP values reported in the previous section are limited either by technology or by regulations.

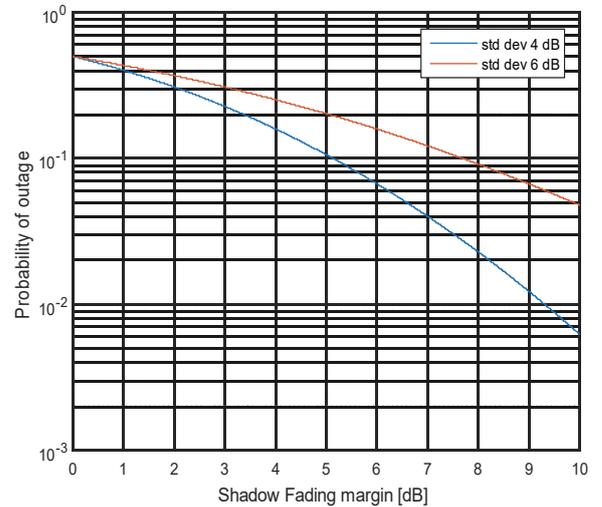

Fig. 3. Lognormal shadowing distribution for standard deviation values recommeded in [16].

Aside from tropospheric fading, which is a well-known effect for satellite communication system operating at mmWaves, there is far less understanding with respect to shadowing and multipath conditions. The reason is that the typical mode of communication at mmWaves is employing highly directional antennas always pointing to the satellite under line-of-sight (LOS). For lower gain or nearly LOS UEs, as the ones encountered in the previous section, there is much less understanding of how the shadowing behaves in various (open, suburban, tree shadowed) environments and elevation angles. An initial assessment based on extending the 3GPP Spatial Channel Model (SCM) to Non Terrestrial Networks (NTN) is provided in [15]. However, further experimental channel campaigns are necessary to calibrate the model for NTN.

---

[2] The 73/83 GHz is the so-called W-band in the satellite terminology. Although this band is not part of the 5G NR FR2, it has been heavily considered by academia and industry for use in mmWave terrestrial radio.

[3] Extrapolating the validity of some of the methods to W-band.

Obviously, in dense urban scenarios and low elevation the LOS is blocked. This may not represent a blocking point as the target complementary coverage for the satellite access should be more the sub-urban or rural scenario. To note that, even employing the lognormal shadowing standard deviation for LOS recommended in the channel modeling document [16] leads in Fig. 3 to significant shadow margins. Furthermore, clutter loss can be estimated via ITU-R Recommendation P.2108 [17] and the UE needs to be outdoors in order to receive from a satellite.

### B. Sensitivity of Downlink Budget

Next we will investigate the sensitivity of the link budget based on typical characteristics of the UE and various channel assumptions. To take a best case approach, we assume that the NGSO satellite always transmits in the downlink at the max power allowed by the on ground PFD. Given this assumption, the choice of NGSO, satellite EIRP and channel bandwidth do not influence the results of the downlink sensitivity plots drawn below. In the following section we will check how realistic this satellite output power is.

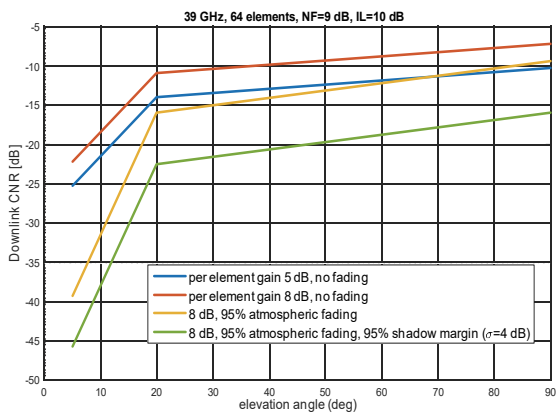

Fig. 4. Downlink CNR versus elevation angle at 39 GHz for various channel assumptions.

Fig. 4 presents the Carrier-to-Noise (CNR) of a 39 GHz downlink against elevation assuming a noise figure (NF) of 9 dB and implementation loss (IL) of 10 dB (see Section III) for various fading assumptions. The UE is assumed to employ 64 elements. Fig. 5 shows the sensitivity of this downlink CNR versus UE number of antenna elements at zenith angle.

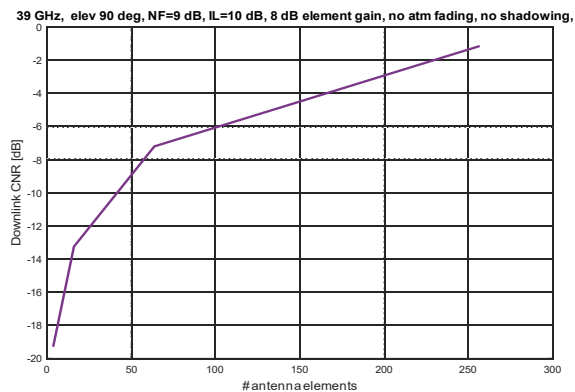

Fig. 5. Downlink CNR versus number of antenna elements at 39 GHz at zenith angle.

Despite the high number of UE antennal elements assumed in the figures above, the downlink budget seems unfeasible for low elevation angles (below say 40 deg.). To avoid the heavy additional fading margins, it is implied that a downlink power control that always adapts to the max PFD limit is in place. Shadowing is however still an open question. Another finding not shown in Fig. 4 and Fig. 5 is that the downlink CNR is much better at Ka-band (20 GHz), where, however, there is no shared allocation with 5G FR2. Notwithstanding the lower fading impact, the higher bands do not profit from the physics: the PFD is stricter, the free space losses higher, the UE elementary antenna gain is the same as in Ka and the technology much more lossy.

### C. Sensitivity of Uplink Budget

Considering now the uplink budget, the following key assumptions have been made: an UE EIRP peak values of 25 dBm (in line with Section III) and a 340 km VLEO satellite with receive G/T=13.5 dB/K that will be motivated in detail in Section V. Along with the 25 dBm peak EIRP, also the (theoretical) upper FCC limit of 43 dBm is considered.

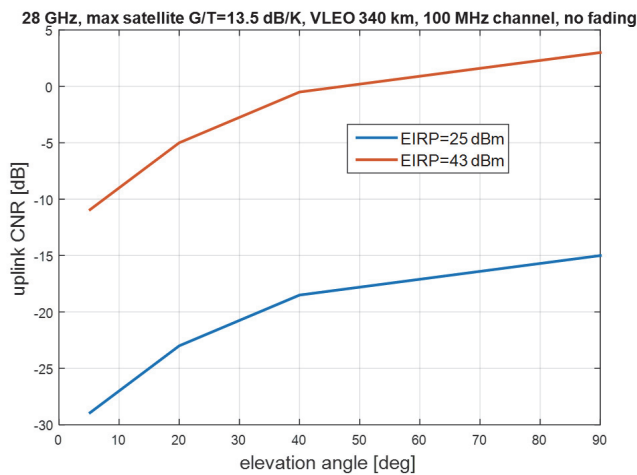

Fig. 6. Uplink CNR versus elevation angle at 28 GHz.

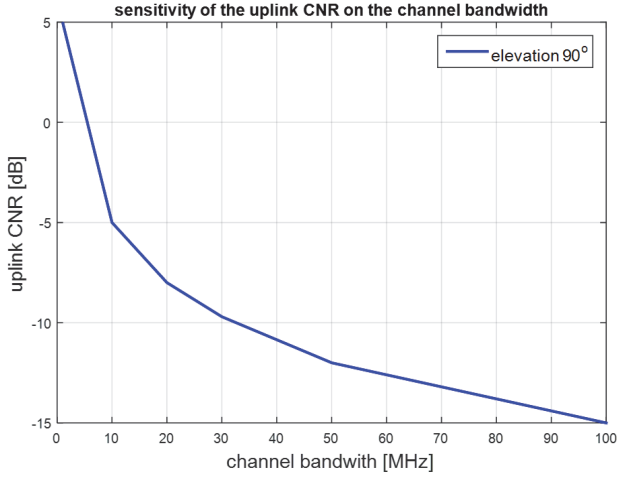

Fig. 7. Sensitivity of uplink CNR versus channel bandwidth at 28 GHz at zenith angle.

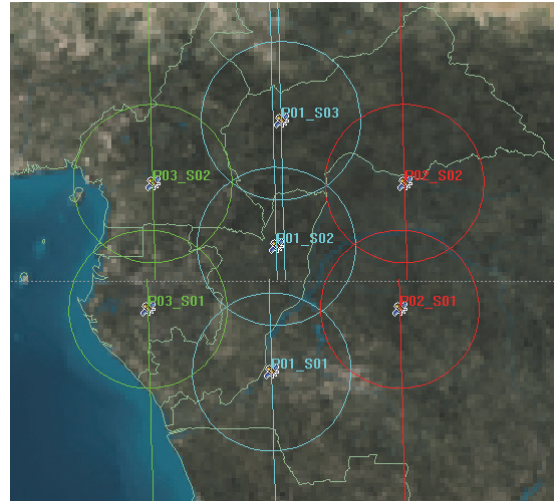

Fig. 8. Seven VLEO satellites in three adjacent planes and their visible Earth projections.

Fig. 6 presents the uplink CNR at 28 GHz against elevation without any fading. The channelization used here is 100 MHz, which although allowed by 5G NR, seems that it cannot be supported by a NGSO link, not even a VLEO. Fig. 7 presents the sensitivity of the uplink CNR on the channel bandwidth for the (best case) of a zenith link. The conclusion is drawn that maybe only an elementary return link from the UE at very high elevation can be expected in such a system scenario with a channel of few MHz.

## V. SPACE SEGMENT

Space segment design depends primarily on the objectives of a satellite operator. As such, this is not the main focus of the paper which is rather aimed at investigating feasibility aspects related to providing 5G mmWave services directly to UEs. On the other hand, it is important to establish the transmit and receive performance of a possible satellite system considering state-of-the-art technologies.

A first analysis has been performed to verify the required size (i.e. number of satellites) of a VLEO constellation for having global continuous coverage from quasi polar orbits. In view of the limited UE capabilities (Section III), the orbit altitude was set to 340 km (reference based on the SpaceX filing for their so-called "Phase 2" satellites). An additional constraint is the minimum elevation. Due to the sharp increase of tropospheric and other attenuation factors at low elevation angles, a minimum elevation angle of 40 deg. is assumed, specifically having a single satellite visible at elevation higher than 40 degrees globally. The constellation is delta-Walker. Fig. 8 shows a cluster of 7 satellites in 3 adjacent planes. The snapshot is taken at an equatorial latitude, which is driving the Right Ascension of Ascending Node (RAAN) spacing between orbit planes. The circles projected on ground represent visible Earth from each satellite at an elevation of 40 deg. These assumptions lead to a minimum of 36 orbit planes with 72 satellites per plane, for a total constellation of 2592 satellites. Despite being a very large constellation, given the deployment rates of currently deployed or planned constellations, this number seems attainable.

A second basic task is sizing the antenna and RF section of the payload to verify feasibility of the technology. The target of the sizing was to achieve a PFD consistent with the SpaceX filing in the transmit direction. Typical satellite telecommunication systems are built in such a way to have a consistent beam size in both transmit (Tx) and receive (Rx), despite the considerable difference in frequency. As in our case the Rx frequency is 28 GHz and Tx frequency is 39 GHz, the diameter of the Rx antenna should have been 39/28 = 1.4 times the diameter of the Tx antenna. In this design exercise, instead it was decided to oversize the Rx antenna to 2 times the Tx diameter in order to support a more favorable link budget in the uplink. The resulting diameters are 40 cm in Rx and 20 cm in Tx based on an hexagonal planar phased array with triangular element lattice. The element lattice is shown in Fig. 9, while TABLE VI. shows the main antenna parameters. Moreover, TABLE VII. and TABLE VIII. show the transmit and receive RF parameters respectively.

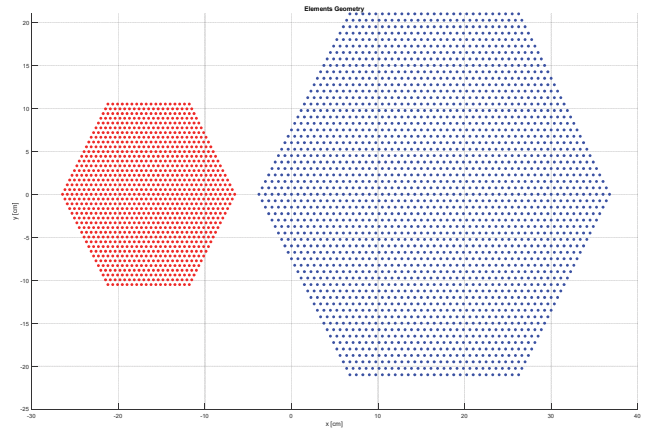

Fig. 9. Satellite Rx array (right) at 28 GHz consisting of 1000 elements and Tx array (left) at 39 GHz consisting of 2000 elements.

TABLE VI. KEY SATELLITE TRANSMIT AND RECEIVE ANTENNA PARAMETERS.

|  | TX DRA |  | RX DRA |  |
|---|---|---|---|---|
| Frequency | 40.00 | GHz | 28.00 | GHz |
| DRA diameter | 0.20 | m | 0.40 | m |
| NGSO Altitude | 340 | km | 340 | km |
| min Elevation | 40 | degrees | 40 | degrees |
| Earth view angle | 71.68 | degrees | 71.68 | degrees |
| Scan Angle | 46.66 | degrees | 46.66 | degrees |
| 3 dB Beamwidth | 1.90 | degrees | 1.36 | degrees |
| 3 dB Beamwidth (scanned beam) | 6.06 | degrees | 4.33 | degrees |
| maximum DRA element Spacing | 0.69 | wavelenghts | 0.69 | wavelenghts |
| maximum DRA element Spacing | 0.52 | cm | 0.74 | cm |
| N elements | 977 |  | 1915 |  |

TABLE VII. KEY SATELLITE ANTENNA TRANSMIT RF CHARACTERISTICS.

| TX DRA | | |
|---|---|---|
| User Forward | | |
| Antenna Gain | 38.09 | dB |
| Max antenna rolloff | -1.00 | dB |
| Output Losses | 1.00 | dB |
| SSPA RF Power | 0.10 | W |
| Total RF power | 99.67 | W |
| Total EIRP | 56.08 | dBW |
| N Beams | 8 | |
| Band per Beam | 0.50 | GHz |
| Total Bandwidth per Satellite | 4.00 | GHz |
| Max slant range | 511.16 | km |
| **EIRP Density** | **-39.94** | **dBW/Hz** |
| **PFD** | **-105.11** | **dBW/m2/MHz** |

TABLE VIII. KEY SATELLITE ANTENNA RECEIVE RF CHARACTERISTICS.

| RX DRA | | |
|---|---|---|
| Antenna Noise Temperature | 300 | K |
| Noise Figure | 2 | dB |
| Input Loss | 0.5 | dB |
| System Noise Figure | 2.5 | dB |
| System Noise Temperature | 533.48 | K |
| G/T | 13.66 | dB/K |

Finally, Fig. 10 shows the 3 dB beam contour in Tx (blue) and Rx (green) at different scan angles and as is evident the difference is reasonably small. In any case, it is expected that state-of-the-art constellation satellites will operate on a per-beam user resource allocation, i.e. beams will be generated and reconfigure over time at a very fast pace, ideally generating one beam centered on the user served at a given time in a time division fashion. As such, no major issue would arise in case of different Tx / Rx beam size.

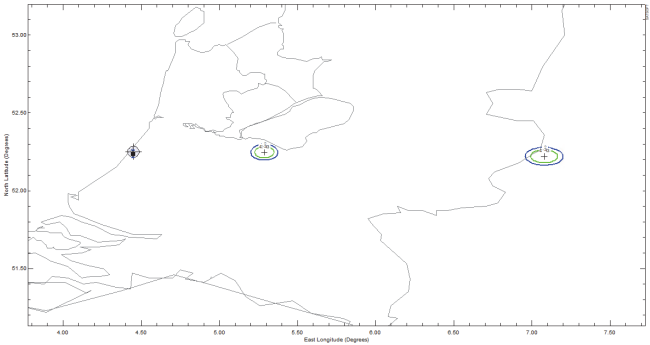

Fig. 10. Transmit and Receive beams at different off-axis angle.

The overall sizing was done assuming that the satellite operator would target rather small satellite platforms and use 4 GHz of available Tx bandwidth. The low RF power per element resulting from the design allows having multiple choices in terms of technology. As an example, a number of factors such as high level of integration, relatively low cost (compared to GaN or even GaS), and current developments led by 5G makes SiGe a very promising option.

## VI. REFINED LINK BUDGET ANALYSIS

Section IV revealed the great challenges of achieving reasonable performance with a system targeting direct access between a NGSO and a typical handheld 5G NR UE at mmWaves. Nevertheless, in this section we identify some more favourable link cases by assuming the space segment described in Section V and a high-end UE from Section III. Specifically, we assume that the UE is of higher class, such a vehicular UE and can, therefore, accommodate more antenna elements and offer a slightly improved noise performance.

TABLE IX. FAVOURABLE DOWNLINK BUDGET AT 39 GHz FROM A VLEO TOWARDS A HIGH PERFORMING UE.

| Q-band | |
|---|---|
| frequency [GHz] | 39 |
| wavelength [m] | 0.007692308 |
| elevation | 40 |
| Max PFD [dBW/m^2/1MHz] | -105 |
| bandwidth [MHz] | 400 |
| Max PFD over bandwidth | -78.98 |
| satellite altitude [km] | 340.00 |
| Link distance [km] | 511.17 |
| max EIRP per carrier [dBW] | 46.18 |
| EIRP density [dBW/Hz] | -39.84 |
| FSL [dB] | 178.43 |
| Noise Figure [dB] | 7 |
| Noise PSD [dBm/Hz] | -167 |
| total noise power [dBm] | -80.98 |
| Atmospheric losses @ 99%/95% [dB] | 5 |
| Shadow fading margin (for std=4dB) @ 95% [dB] | 0 |
| number of terminal antenna elements | 256 |
| antenna patch single dimension [cm] | 6.15 |
| gain per element [dBi] | 8 |
| Rx terminal antenna gain [dBi] | 32.08 |
| total receiver noise temp [dBK] | 31.60 |
| G/T [dB/K] | 0.48 |
| received carrier power after user antenna [dBm] | -75.17 |
| implementation loss [dB] | 7.00 |
| C/N [dB] | -1.19 |
| Spectral efficiency with 5G NR ModCod [bps/Hz] | 0.5 |
| Data Rate [Mbps] | 200 |

TABLE IX. shows in detail such a downlink budget at 39 GHz, adopting the Tx characteristics of the VLEO of the previous section communicating with a UE equipped with 256 antenna elements. In terms of channelization, the NR waveform allows for 50, 100, 200 and 400 MHz. In TABLE IX. , 400 MHz has been selected in order to match the VLEO EIRP density. For all channel effects, a flat margin of 5 dB has been injected in the link budget. Even under these favourable assumptions, the UE can barely close the link with a robust 5G NR modulation and coding scheme of high spectral efficiency. Note that TABLE IX. disregards the effect of co-

channel interference from other satellites in the networks. This is done as it is typically assumed that UEs will be sparse compared to the satellite coverage area (see Fig. 10) and that the interference can be handled by means of radio resource management and beam management. Moreover, the slightly negative downlink CNR testifies to the fact that the link is much more power limited rather than interference limited.

If one translates this link performance into system capacity, it means that there will be (200 Mbps) x (1 carrier/beam) x (8 beams/satellite) = 1.6 Gbps per satellite or (2592 satellites) x (1.6 Gbps / satellite) = 4.1 Tbps for the whole constellations. Of course, these numbers are much lower than the ones estimated for other constellations [2], but this is simply due to the less performing UEs.

TABLE X.    FAVOURABLE UPLINK BUDGET AT 28 GHZ FROM A HIGH PERFORMING UE TOWARDS A VLEO.

**Ka-band**

| | |
|---|---|
| frequency [GHz] | 28 |
| wavelength [m] | 0.0107 |
| elevation | 40 |
| bandwidth [MHz] | 1 |
| satellite altitude [km] | 340.00 |
| max EIRP per carrier [dBm] | 29.00 |
| Link distance [km] | 511.17 |
| FSL [dB] | 175.56 |
| Atmospheric losses @ 99%/95% [dB] | 5 |
| Shadow fading margin (for std=4dB) @ 95% [dB] | 0 |
| satellite G/T [dB/K] | 13.50 |
| Boltzmann | -228.60 |
| implementation loss [dB] | 0.00 |
| C/N[dB] | 0.54 |
| Spectral efficiency with 5G NR ModCod [bps/Hz] | 0.66 |
| Data Rate [Mbps] | 0.66 |

In the uplink direction, a higher class UE is selected that can output an EIRP of 29 dBm towards the VLEO at 28 GHz. Such an EIRP seems to be feasible for a vehicular UE, but even a handheld UE in the boresight region [19]. Due to the very weak uplink, it is necessary to consider the minimum channel bandwidth allowed by NR in FR2. The minimum allowed NR channel needs to be 1 Bandwidth part (BWP), that is at least 1 Resource Block (RB). Then the minimum RB in FR2 is 12 sub-carriers x 60 kHz = 720 kHz minimum channel bandwidth. We have rounded this to 1 MHz for possible overhead and guard bands. TABLE X. presents the resulting (favourable) uplink performance, however with only 5 dB of a fading margin. Due to the minimum bandwidth assumption, the low spectral efficiency leads to very low data rate, maybe just enough to sustain some signalling in the return link. For the same reasons as for the downlink, co-channel interference from within the system is not considered on the uplink in this preliminary analysis.

### VII. BEAM MANAGEMENT & SYSTEM ASPECTS

The link budget analysis carried out in the previous section was assuming that the pointing between the UE and the satellite is given. However, 5G NR at mmWaves relies on a new paradigm compared to 4G as initial access, control and data communications all rely on directional beams. This is accomplished by implementing the so-called *Beam management* concept, which is a set of Layer 1/2 procedures to acquire and maintain a set of BS and/or UE beams [5]. It supports analog beam-forming at both the BS and the UE side. Beam management involves covering an angular sector by sweeping analog beams within that sector, measuring their quality, reporting the best beams, selecting one or few of them, switching to another beam if it gets higher quality, or recovering a new beam if the current one is blocked. This process relies on an extensive signalling framework, which involves in the downlink the Signal Synchronization Block (SSB) and the Channel State Information – Reference Signal (CSI-RS). For uplink beam management, the Sounding Reference Signal (SRS) is used [5], [20]. Although it is outside the scope of the present paper, this section attempts to raise the issue of whether these reference signals can be applied over a NGSO network latency, even if its a VLEO one. This definitely deserves further research.

### VIII. CONCLUSIONS & FUTURE RESEARCH

The paper carries out a preliminary feasibility analysis on the possibility of establishing direct broadband communications between a NGSO satellite (with emphasis on VLEO) and a 5G NR enabled (handheld or vehicular) UE in the mmWave range (FR2). The paper takes a regulatory, UE characteristics, space segment, link budget and system point of view. Having identified the major challenges which is the use of TDD for 5G FR2 and the very weak link budgets, the paper proposes a number of further investigations to complete the study:

- Improve the understanding of the regulatory landscape, how to share bands between FSS and 5G NR, possibility for implementing TDD in FSS.
- Characterize the land mobile satellite channel in terms of shadowing, clutter and multipath low gain UEs at mmWaves[4].
- 5G NR Beam management provides a wealth of signalling tools for analog UE beamforming. The applicability of this framework to FSS even for carrying out basic pointing procedures is unclear at this time.
- Technology developments on how can satellite industry profit from commercial low cost antenna developments for the mass market that are currently in full speed ahead for 5G.

ACKNOWLEDGMENT

The authors wish to thank their colleagues from ESA's Telecommunications and Integrated Applications (TIA) directorate for supporting this work, in particular the ARTES Future Preparation Programme https://artes.esa.int/future-preparation. The views and opinions expressed in this paper are by no way representative of the views of the Agency.

---

[4] In this respect, ESA has already issued an ITT under the ARTES AT programme, namely 'Q-band channel model for mobile terminals' https://artes.esa.int/funding/qband-channel-model-mobile-terminals-artes-3b038-0